\def\year{2023}\relax
\documentclass[letterpaper]{article}
\usepackage{aaai}
\usepackage{times}
\usepackage{helvet}
\usepackage{courier}
\newcommand{\proposed}{SSL-Auth}
\usepackage{graphicx}
\usepackage{stfloats}
\usepackage{amsmath}
\usepackage{multirow}
\usepackage{subfigure}
\usepackage{algorithm}  
\usepackage{algpseudocode}
\usepackage{url}
\usepackage{booktabs}
\usepackage{colortbl}
\usepackage{xcolor}
\usepackage{makecell}
\nocopyright

\begin{document}
%
\title{SSL-Auth: An Authentication Framework by Fragile Watermarking for Pre-trained Encoders in Self-supervised Learning}
\author{Xiaobei Li\textsuperscript{\rm 1}, 
    Changchun Yin\textsuperscript{\rm 1},
    Liyue Zhu\textsuperscript{\rm 1},
    Xiaogang Xu,\textsuperscript{\rm 2},
    Liming Fang\textsuperscript{\rm 1},
    Run Wang\textsuperscript{\rm 3}, 
    Chenhao Lin\textsuperscript{\rm 4}\\
    \textsuperscript{\rm 1} Nanjing University of Aeronautics and Astronautics \\
    \textsuperscript{\rm 2} Zhejiang Lab \\
    \textsuperscript{\rm 3} Wuhan University \\
    \textsuperscript{\rm 4} Xi'an Jiaotong University \\
    \{xiaobei\_li\_am, ycc0801, sx2316068\}@nuaa.edu.cn, xgxu@zhejianglab.com, \\fangliming@nuaa.edu.cn, wangrun@whu.edu.cn, linchenhao@xjtu.edu.cn
}
\maketitle
\begin{abstract}
\begin{quote}
Self-supervised learning (SSL), a paradigm harnessing unlabeled datasets to train robust encoders, has recently witnessed substantial success. These encoders serve as pivotal feature extractors for downstream tasks, demanding significant computational resources. 
Nevertheless, recent studies have shed light on vulnerabilities in pre-trained encoders, including backdoor and adversarial threats. Safeguarding the intellectual property of encoder trainers and ensuring the trustworthiness of deployed encoders pose notable challenges in SSL.
To bridge these gaps, we introduce SSL-Auth, the first authentication framework designed explicitly for pre-trained encoders. SSL-Auth leverages selected key samples and employs a well-trained generative network to reconstruct watermark information, thus affirming the integrity of the encoder without compromising its performance. By comparing the reconstruction outcomes of the key samples, we can identify any malicious alterations. 
Comprehensive evaluations conducted on a range of encoders and diverse downstream tasks demonstrate the effectiveness of our proposed SSL-Auth.   
\end{quote}
\end{abstract}
\section{Introduction}
The most common method in machine learning is supervised learning, which relies on a large amount of high-quality labeled data. However, the acquisition of such high-quality labels often proves to be a daunting task, necessitating the expertise of experienced human annotators, which is both expensive and time-intensive. Self-supervised learning (SSL) \cite{devlin2018bert,hjelm2018learning,he2020momentum,chen2020improved} offers a compelling solution by advocating for the pre-training of encoders on unlabeled data. This approach effectively circumvents the label scarcity inherent in supervised learning, yielding commendable performance across various downstream tasks.

Nonetheless, the process of collecting and training data for SSL encoders is not without its own challenges. Notably, robust encoders benefit immensely from larger datasets and more powerful computational resources, rendering the endeavor financially demanding. The expense associated with training a high-performance encoder via SSL proves to be a significant barrier for individual practitioners, thereby relegating this task to companies with ample computational resources, predominantly for commercial applications.

\begin{figure}[!t]
	 \centering
\includegraphics[width=0.8\linewidth]{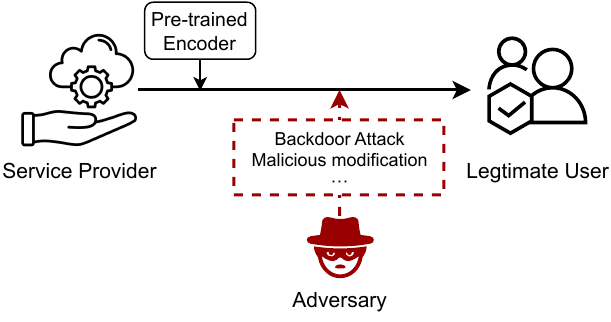}
\caption{Threats to self-supervised learning. When the encoder is sent by the service provider to legitimate users, it may be maliciously modified (e.g. embedding backdoors) by illegal attackers, thus posing a threat to user security. }
\label{intro-1}
\end{figure}

This paradigm also exposes encoders to various potential threats, including stealing attacks \cite{dziedzic2022difficulty,liu2022stolenencoder,sha2023can} and backdoor attacks \cite{carlini2021poisoning,saha2022backdoor,li2022demystifying,xue2022estas}. As illustrated in Figure \ref{intro-1}, attackers may maliciously tamper with the encoder during its delivery to users, thereby posing a direct threat to their security and rights. Consequently, users must conduct a thorough integrity check on the received encoder before deploying it, ensuring it remains untainted.

To verify model integrity, a common approach is model watermarking, which can be broadly categorized into two types based on their objectives: robust watermarks \cite{uchida2017embedding,adi2018turning,darvish2019deepsigns,chen2019deepmarks,zhong2020protecting,peng2022fingerprinting} and fragile watermarks \cite{guan2020reversible,botta2021neunac,zhu2021fragile,yin2022neural,lao2022deepauth}. Historically, research in model watermarking primarily centered around specific task classifiers, involving the embedding of meticulously crafted key samples and specific category labels as watermark information into the model. Robust watermarks aim for the model to retain the watermark information even after modification, whereas fragile watermarks necessitate the model to fail verification even following minor adjustments. By incorporating fragile watermarks into the model, one can ascertain if any alterations have been made. A low success rate in watermark verification implies that the model has been tampered with.

Given the adaptability of SSL encoders across diverse downstream tasks, conventional model watermarking approaches encounter challenges, as they are tailored to specific tasks. Furthermore, current research on encoder watermarking, to the best of our knowledge, predominantly revolves around robust watermarking schemes, aiming to embed watermarks that resist modifications or stealing attacks on encoders \cite{cong2022sslguard,dziedzic2022dataset,zhang2023awencoder,wu2022watermarking,lv2022ssl} to validate encoder ownership. The domain of SSL encoder integrity verification remains relatively unexplored.

\begin{figure}[!t]
	 \centering
\includegraphics[width=0.9\linewidth]{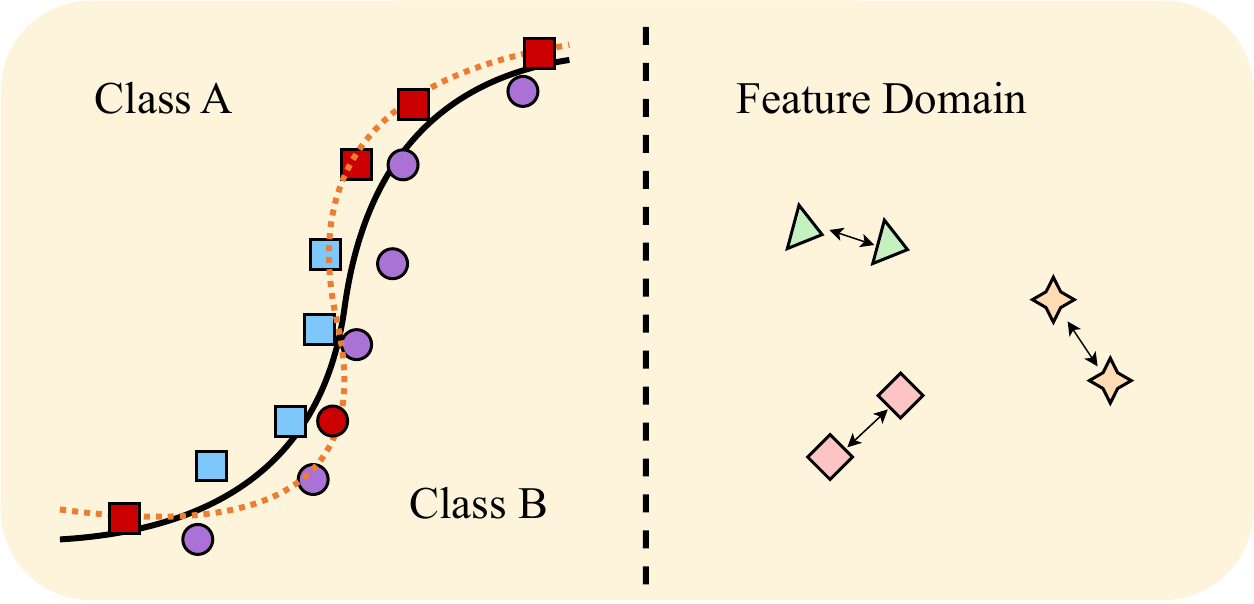}
\caption{The role of key samples in SL fragile watermarks. In SL, changes in classification boundaries can be used to indicate whether the model has been tampered with, while in SSL, only the offset of feature vectors can be used to indicate. }
\label{intro-2}
\end{figure}

In this paper, we introduce \textbf{\proposed}, a pioneering extension of the concept of fragile watermarking to SSL encoders.
Similar to fragile watermarks in supervised learning (SL), our approach assesses encoder integrity by scrutinizing changes in the output of the encoder on key samples. It is important to note that there exists a notable distinction between SL and SSL fragile watermarks. 

In SL, the process entails identifying samples that are situated in close proximity to the model's classification boundary, or embedding specific samples near said boundary. As illustrated in Figure \ref{intro-2}, when the model undergoes tampering, alterations in the label of these samples serve as indicators of subtle shifts in the classification boundary. In contrast, the output of an SSL encoder does not manifest as a categorical label within a defined range, but rather as a determined feature vector. Consequently, when tampering occurs with the encoder, detection hinges on discerning the offset within the feature vector.

\proposed~employs random samples with the same distribution as the pre-training dataset of the encoder as the requisite key samples for validation. We subsequently compare the similarity of the feature vectors generated by these key samples between the original encoder and the suspicious encoder. Regarding similarity assessment, we initially directly compare the feature vectors to compute their similarity within the feature space. To address potentially more formidable attacks, such as adversaries training malevolent encoders to mimic the output of the original encoder for key samples, we further transform the feature vectors into the image domain, aiming to accentuate their disparities. 

In summary, our main contributions are as follows:

\begin{itemize}
\item We propose \proposed, the first authentication framework utilizing fragile watermarks to verify the integrity of SSL encoders. Compared to existing work, \proposed~is both simple and convenient in implementation, and can effectively detect tampered encoders.
\item We propose two novel verification schemes to detect whether the encoder has been maliciously tampered with. Both schemes can effectively verify the integrity of the encoder even when facing strong adversary. 
\item We conduct extensive experiments to evaluate \proposed~on various settings. The results show that \proposed~can successfully distinguish whether the encoder has been modified even if the modified operations have minimal impact on the model.

\end{itemize}

\section{Related Works}
\label{section:Related Works}

\subsection{Self-supervised Learning}
Self-supervised learning (SSL), an approach leveraging the inherent structure of data for learning without the need for labeled supervision signals, has emerged as a powerful technique. In the realm of computer vision, SSL is instrumental for feature learning in images and videos. This is achieved, for instance, by creating self-supervised tasks through operations like rotation and occlusion on images, and subsequently learning the art of extracting valuable features through training.

In this paper, we consider a popular SSL algorithm, SimCLR~\cite{chen2020simple}, which is based on the idea of contrastive learning and learn how to compare samples to extract useful features. It has delivered notable results in the domain of image representation learning, serving as an influential reference and a source of inspiration for the development of SSL methods in diverse fields.

\subsection{Model Watermarking}
In supervised learning, protecting model copyright often involves employing watermarking methods, which typically fall into two categories. One approach relies on backdoor methods \cite{adi2018turning,zhong2020protecting}, while the other directly embeds the watermark into the latent or feature space of the neural network, known as feature-based methods \cite{uchida2017embedding,darvish2019deepsigns,chen2019deepmarks}.

Detecting whether a protected model has been illegally modified during model release is usually achieved through fragile or reversible watermarks. A ``fragile'' watermark implies that even slight interference with the protected model will result in a change in the watermark \cite{zhu2021fragile,yin2022neural,lao2022deepauth}. On the other hand, a ``reversible'' watermark indicates that the protected model can be restored to its original state using the correct key, while any modified version cannot \cite{guan2020reversible,botta2021neunac}.

Within the realm of SSL, there have been some works on protecting model copyright, which attempt to \textbf{inject} a watermark into the pre-trained encoder that can resist transformation attacks (such as fine-tuning, pruning, and overwriting watermarks) and stealing attacks. Noteworthy examples include SSLGuard \cite{cong2022sslguard}, which employs shadow datasets to simulate stealing attacks and trains a decoder to distinguish protected models. Additionally, SSL-DI \cite{dziedzic2022dataset} refrains from modifying the encoder and instead utilizes data density in the representation domain as a boundary, utilizing the encoder's predicted train-test discrepancy as the evaluation metric. However, the proposed \proposed~aims to \textbf{extract} the internal information of the encoder.
\begin{figure*}[!t]
	 \centering
\includegraphics[width=0.75\linewidth]{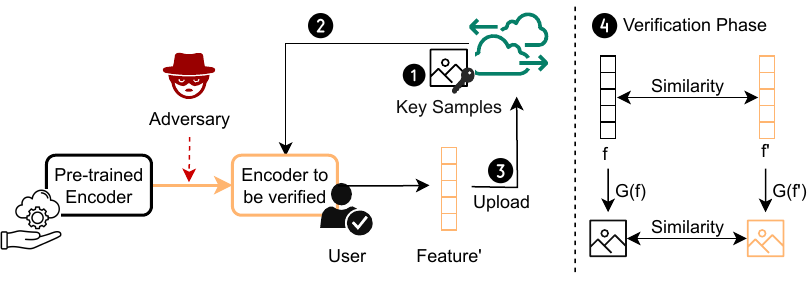}
\caption{Proposed \proposed~mainly consists of 4 steps. First, select key samples and save them to the trusted third-party platform (Step 1). Then, send key samples to the user when receiving the integrity verification request (Step 2). After receiving the feature vector upload from the user (Step 3), verification (Step 4) is then performed.}
\label{method-1}
\end{figure*}

\section{Threat Model}
\label{3}
In this work, we focus on pre-trained encoders with image classification as the downstream task. Referring to the settings in DeepAuth \cite{lao2022deepauth}, as shown in Figure \ref{method-1}, after the training party delivers the encoder to the user, integrity verification is performed through a trusted third-party platform. This trusted third party can query the original encoder and store the key samples used for verification on their server. Upon receiving a user's request for integrity verification, the corresponding key samples are transmitted to the user.

Then, we introduce the adversary’s goal and outline the varying levels of background knowledge they may possess. The extent of an attacker's background knowledge significantly influences their capabilities. 

\textbf{Adversary’s Goal.} Following previous work \cite{botta2021neunac,lao2022deepauth,yin2022neural}, the adversary's goal is to subvert the encoder, potentially by implanting a backdoor. Such tampering can lead to a degradation in the encoder's performance, subsequently impacting downstream tasks. Nevertheless, any alterations should be as inconspicuous as possible; excessive changes would likely draw immediate attention and render the attack futile.

\textbf{Adversary’s Background Knowledge.}We categorize the adversary's background knowledge into two dimensions: knowledge of the pre-trained encoder and knowledge of the key samples used for integrity verification.

For the knowledge of the pre-trained encoder, like DeepAuth \cite{lao2022deepauth}, we assume that the adversary can obtain the complete encoder. Building on this premise, we further assume that the adversary is privy to the pre-trained dataset of the encoder. This empowers the adversary to make alterations that preserve the overall performance of the encoder to the greatest extent possible.

For the knowledge of key samples, we have considered a more stringent scenario based on \cite{lao2022deepauth} and divided adversaries into two types according to their knowledge of key samples, denoted as $\mathcal{A}$ and $\mathcal{B}$. Both possess identical knowledge of the pre-trained encoder. However, Adversary $\mathcal{A}$ lacks any information about key samples, whereas Adversary $\mathcal{B}$ is equipped with knowledge of these samples. This distinction grants Adversary $\mathcal{B}$ the ability to camouflage any tampering with the encoder, effectively deceiving the integrity testing process. 

\section{Proposed Methods}
The result of the self-supervised training process yields an encoder, denoted as $F$, such that for a given input $\mathbf{x}$, we can derive the corresponding feature vector $f=F(\mathbf{x})$. The adversary's goal is to surreptitiously manipulate the original encoder to produce an illicitly modified version $F'$. Depending on the adversary's varying background knowledge, our framework incorporates distinct schemes to assess the integrity of the encoder. 

\textbf{Overview.}
Figure \ref{method-1} summarizes the entire process of \proposed: 1) the framework commences with the selection of key samples, which are then uploaded to the trusted third-party platform for the purpose of encoder integrity verification; 2) upon receiving an integrity verification request from the user, the platform transmits the relevant key sample to the user; 3) the user inputs the key sample into the encoder to be verified and subsequently uploads the corresponding output back to the platform; 4) finally, the platform compares the differences of the outputs between the original encoder and the encoder to be validated, and returns the authentication success rate and the overall similarity of key samples to the user.

\subsection{Key Samples} 
The key samples $K$ employed for encoder integrity verification are selected at random. This approach stems from the fact that the output of an SSL pre-trained encoder does not yield precise category labels, obviating the need for specific samples to discern potential tampering. We have discussed the impact of different selected K in \ref{A2}.

It is crucial to note that the pre-training data $\mathcal{D}_t$ utilized by the encoder lacks labels, and the encoder $F$ lacks a distinct classification boundary—a stark departure from the landscape of supervised learning. In the realm of supervised learning, discerning model performance often entails generating or selecting key samples proximate to the classification boundary, enabling the detection of subtle shifts in the said boundary. This is because models in supervised learning typically output categorical labels. Consequently, if a model undergoes slight modifications, samples not residing on the classification boundary are likely to remain within the original category.

Through vigilant monitoring of alterations in output features, we can effectively identify encoder modifications. Even minute adjustments to the encoder can yield notable disparities in encoding features. Therefore, our approach hinges on leveraging the intrinsic properties of the encoder, using features from the key samples to uncover any potential tampering, rather than relying on explicit category labels. 

Learned from \cite{dziedzic2022dataset}, the encoder and its derivatives, such as stolen copies and modified copies, exhibit more similar behavior on the pre-training dataset. It can be represented as $\mathcal{E}(F'(\mathcal{D}_t)) > \mathcal{E}(F'(\mathcal{D}_n))$, where $\mathcal{E}$ is a density estimator trained by a dataset with the same distribution as $P_{\mathcal{D}_t}$, $P_{\mathcal{D}_t}$ denotes the distribution of $\mathcal{D}_t$ and $\mathcal{D}_n$ is a normal dataset that $P_{\mathcal{D}_t} \neq P_{\mathcal{D}_n}$. 

Therefore, we select key samples from a distribution that is different from $P_{\mathcal{D}_t}$, which leads to increased uncertainty on the outputs when the encoder is modified. 

\subsection{Verification Phase}
\label{4.2}
\textbf{Direct Comparison (DCmp).} 
In a high-dimensional space, two randomly selected vectors tend to be nearly orthogonal, resulting in their cosine similarity being clustered around 0. Conversely, if the cosine similarity between two vectors significantly exceeds 0 or approaches 1, this indicates a close relationship between them. As such, cosine similarity emerges as a valuable metric for comparing the features extracted by the original encoder and those extracted by the suspicious encoder, ultimately serving as a means to demonstrate the trustworthiness of the latter:
\begin{equation}
{\mathcal{L}_{\cos}}(a,b)=\frac{\vec{a} \cdot \vec{b}}{|\vec{a}| \cdot|\vec{b}|}. 
\end{equation}

The value of cosine similarity serves as a metric that reflects the degree of correlation between vectors. When dealing with randomly selected key samples $\mathbf{K}$, the trusted third-party platform can query the original encoder to obtain $f_{K}=F(\mathbf{K})$, compare it with the user-uploaded $f_{K}'=F'(\mathbf{K})$, and calculate cosine similarity directly. The resulting calculation falls within the interval of 0 to 1, representing the similarity of the encoder in terms of the correlation between the feature vectors corresponding to the key samples. A cosine similarity of 1 across all feature pairs signifies that the encoder has not undergone any illicit tampering. To account for potential accuracy loss in the feature vectors uploaded by users, which may impact the calculation results, a cosine similarity very close to 1 can also be deemed indicative of a successful sample verification.
Only if all samples in the key samples pass the verification, then \proposed~considers the watermark verification successful.

\textbf{Transformation (Trans).} 
If the adversary obtains key samples, they can potentially manipulate the encoder in a manner that conceals their tampering. Their goal is to ensure that the modified encoder produces feature vectors for key samples that closely resemble the output of the original encoder. By achieving this, the integrity verification process can be misled. To address this issue, our approach seeks to magnify the initially subtle discrepancies by converting the feature vectors into alternate dimensions or domains. 

One idea is to convert feature vectors to higher or lower dimensions while preserving their vector form, subsequently employing cosine similarity for comparison. Another idea entails the restoration of feature vectors into images, enabling a comparison of the similarity shifts between the images to determine whether tampering has occurred.

This conversion process from feature vectors to images is facilitated by a trusted third-party platform, which initiates a generator as a convenient tool. In this paper, we leverage a Generative Adversarial Network (GAN) \cite{goodfellow2014generative} comprised of a generator model, denoted as $G$, to reconstruct the feature vectors into images. $G$ can generate complex outputs from high-dimensional vectors. This ability makes $G$ a powerful tool for exploring and visualizing high-dimensional data structures. During the generation process, $G$ introduced randomness. This means that even for very similar inputs, the outputs may vary. Note that we only use the output of $G$ to compare the differences between vectors, without considering the quality and meaning of the reconstructed image.

By training the generator to minimize the similarity between the reconstructed images from feature vectors and the key samples, we can make small changes in the input vectors more pronounced in the generated images. This is because the generator learns to make the outputs dissimilar to the key samples. Images distinct to key samples do not necessarily have similar features, resulting in relatively lower similarity between the outputs of the original feature vectors and the changed feature. The training goal of the generator can be represented by the following formula: 
\begin{equation}
\min_{G} \mathcal{L} = sim(G(F(\mathbf{K})), \mathbf{K}).
\end{equation}

This transformation shifts the similarity assessment from one of the feature vectors to that of images. Numerous loss functions can be employed to quantify image similarity, including ${\mathcal{L}_1}$ loss (Mean Absolute Error, MAE) and ${\mathcal{L}_2}$ loss (Mean Squared Error, MSE), among others. For tasks involving image restoration, if the detection of outliers is pertinent, MSE may be a preferable choice over MAE. However, it's worth noting that MSE yields values within the range of 0 to $\infty$, which cannot be interpreted as a percentage to quantify the reconstruction effect.

Here, we use SSIM Loss~\cite{wang2004image} to measures the similarity between two images in terms of structure, color, and brightness. It consists of three loss function: luminance loss $l(\mathbf{x}, \mathbf{y})=\frac{2 \mu_{ x} \mu_{y}+c_{1}}{\mu_{x}^{2}+\mu_{y}^{2}+c_{1}}$, contrast loss $c(\mathbf{x}, \mathbf{y})=\frac{2 \sigma_{x} \sigma_{y}+c_{2}}{\sigma_{x}^{2}+\sigma_{y}^{2}+c_{2}}$, and structure loss $s(\mathbf{x}, \mathbf{y})=\frac{\sigma_{x y}+c_{3}}{\sigma_{x} \sigma_{y}+c_{3}}$.
Multiplying these three loss functions yields the expression for SSIM loss:
\begin{equation}
\begin{aligned}
\mathcal{L}_{SSIM}(\mathbf{x}, \mathbf{y}) &= l(\mathbf{x}, \mathbf{y})\cdot c(\mathbf{x}, \mathbf{y})\cdot s(\mathbf{x}, \mathbf{y}) \\
\end{aligned}
\end{equation}
\section{Evaluation}
In this section, we first introduce the experimental setup in Section \ref{5.1}. Then, we show the performance of \proposed~on common model modification operations in Section \ref{5.2}. Next, we evaluated the effectiveness of \proposed~in the face of stronger adversaries in Section \ref{5.3}. We also evaluate the performance on precision loss in Section \ref{5.4}. Finally, we compared the fidelity of the existing work in Section \ref{5.5}.

\subsection{Experimental Setup} 
\label{5.1}
\textbf{Datasets.} We use the following 5 public datasets to conduct our experiments.

\begin{itemize}
    \item {\bf ImageNet~\cite{russakovsky2015imagenet}.} Its subset ILSVRC2012 is being widely used, which includes $1000$ categories, 1.2 million training set images and $150000$ test set images. 
    \item {\bf CIFAR-10~\cite{krizhevsky2009learning}.} This is a small dataset for identifying ubiquitous objects, consisting of a total of 10 categories, with an image size of $32 \times 32 \times 3$. 
    \item {\bf STL-10~\cite{coates2011analysis}.} The images in STL10 are from ImageNet, with a total of 113000 images with a size of $96 \times 96 \times 3$, including $100000$ unlabeled images. \footnote{\url{https://cs.stanford.edu/~acoates/stl10/}} 
    \item {\bf GTSRB~\cite{stallkamp2012man}.} German Traffic Sign Recognition Benchmark (GTSRB) is used for identifying traffic signs, containing 43 types of traffic signs. 
    \item {\bf SVHN~\cite{netzer2011reading}.} This dataset is collected from the house number in Google Street View images, consisting of ten classes, with an image size of $32 \times 32 \times 3$. 
    
\end{itemize}

In our experiments, we leverage both ImageNet and CIFAR-10 as pre-training datasets for the encoder. Additionally, we use STL-10, GTSRB, and SVHN as the downstream datasets for evaluation.

\textbf{Pre-trained Encoders and Implementation Details.}
In our experiments, we employ a pre-trained encoder from the real world as the protected encoder when utilizing ImageNet as the pre-training dataset. Specifically, we acquire the encoder checkpoints from the official SimCLR repository\footnote{\url{https://github.com/google-research/simclr}}. When CIFAR-10 serves as the pre-training dataset, we conduct the training of SimCLR encoders on one NVIDIA GeForce RTX 3090 GPU. All encoders utilized are based on ResNet-50 architecture.

For the generative model $G$, we opt for PGAN~\cite{karras2017progressive} as the default choice and trained for 1000 epochs. For key samples, we randomly selected 100 samples from the pre-training datasets ImageNet and CIFAR-10. Then, we swapped the distributions of their corresponding key samples. This means that during the validation process, we used key samples from CIFAR-10 for the ImageNet pre-trained encoder, and key samples from ImageNet for the CIFAR-10 pre-trained encoder.

\textbf{Tolerance.}
In the fragile watermark task, we hope to detect whether the feature vectors of key samples have changed.
Given the potential for precision loss when users upload feature vectors, even minor changes in the vectors might lead to a slight reduction in similarity. Consequently, to safeguard the validation process from being unduly affected by precision loss, we have established a stringent similarity threshold $\epsilon$ of \textbf{99.5\%}.
When the similarity between two vectors is greater than $\epsilon$, it can be considered that the sample has passed validation. This fault-tolerant constraint allows for an error range, which can ensure the reliability of validation to a certain extent while reducing the possibility of validation failure caused by precision loss.

\textbf{Evaluation Metrics.}
We use agreement and authentication success rates to evaluate the fragile watermark’s performance. The function $\mathcal{S}(\cdot)$ we use to calculate sample similarity varies depending on the validation method we choose. When we compare vectors directly (DCmp) for validation, $\mathcal{S}(f_{K},f_{K}')=\mathcal{L}_{cos}(f_{K},f_{K}')$; When we convert feature vectors into the image domain (Trans) for comparison, $\mathcal{S}(f_{K},f_{K}')=\mathcal{L}_{SSIM}(G(f_{K}), G(f_{K}'))$.
\begin{itemize}
    \item {\bf Agreement.}
    For each sample ${\bf k_i}$ in key samples $\mathbf{K}$, calculate its similarity to the output of the original encoder and the encoder to be verified. 
    \begin{equation}
    \text {Agreement}_i=\mathcal{S}(F(\mathbf{k_{i}}),F'(\mathbf{k_{i}})). 
    \end{equation}
    $\text {Agreement}=\mathcal{S}(F(\mathbf{K}), F'(\mathbf{K}))$ can be used to reflect the degree to which the encoder has been modified. Note that we do not use the agreement as a metric to determine whether the encoder has passed validation. Agreement is only used to reflect the overall similarity of the encoder. 
    
    \item {\bf Authentication Success Rate.}
    When the similarity corresponding to $k_i$ is greater than $\epsilon$, we believe that this sample has passed validation. 
    \begin{equation}
    \text {Auth Success Rate}=\frac{\sum_{i=1}^{n} \text {Agreement}_i>=\epsilon}{n}. 
    \end{equation}
    The fragile watermark expects a low authentication success rate when the encoder has been modified. When the authentication success rate reaches 100\%, the verification is considered successful. 
        
\end{itemize}

\subsection{Performance on Common Model Modification Operations}
\label{5.2}
The goal of adopting \proposed~is to ensure that any illegal tampering of the encoder can be detected. To achieve this goal, we evaluated the performance of detecting encoder modifications through fine-tuning, pruning, and backdoor attacks. Although fine-tuning and pruning sometimes are not attacks, the encoder after processing still maintains similar performance, which better demonstrates the fragility of the watermark. Our experiment results show that even slight modifications to the protected encoder cannot pass the verification, thus detecting potential integrity destruction.

Here, we also compared the performance of two verification schemes mentioned in Section \ref{4.2} for detecting whether the encoder has been tampered with. The transformation scheme we adopted is the method of converting the feature vectors into images with a size of $32\times32$ through $G$.

\begin{figure}[t]
 \centering
\includegraphics[width=1\linewidth]{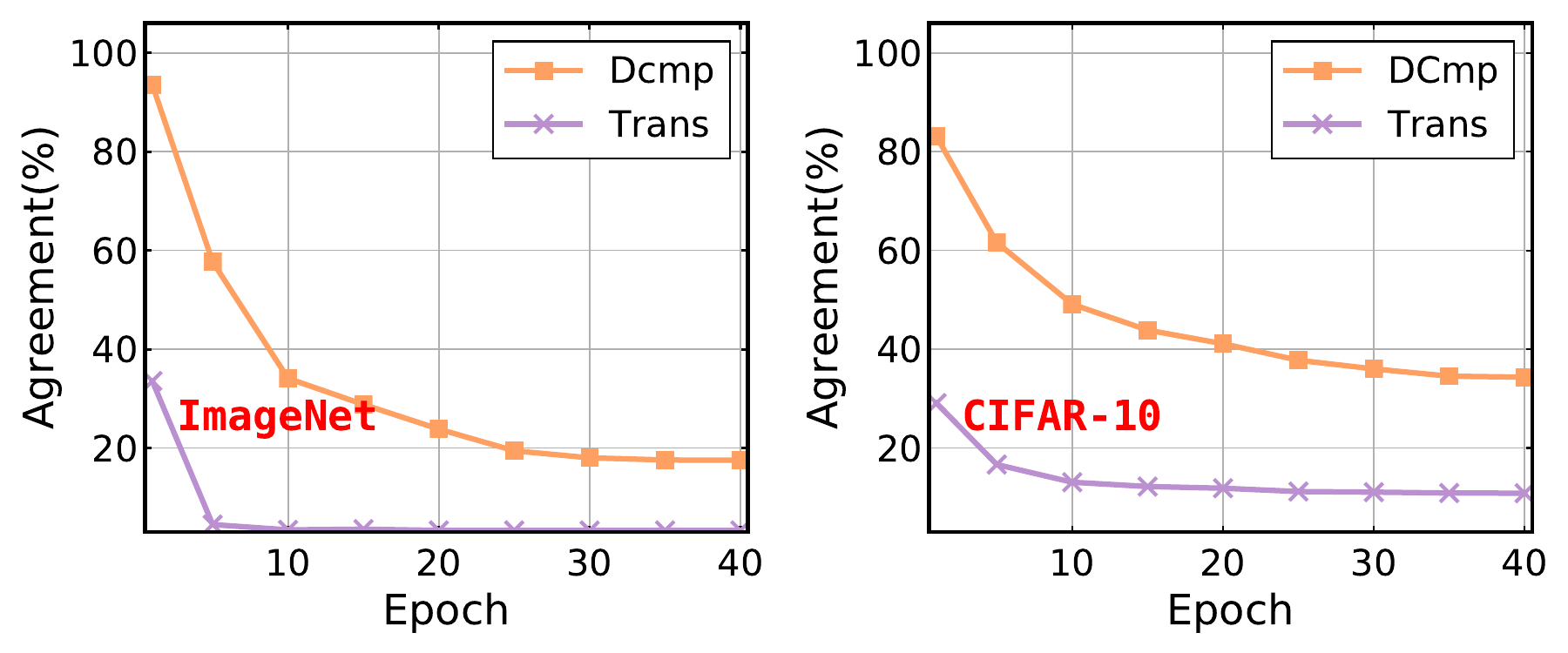}
\caption{Agreement on encoder fine-tuning. As the epoch increases, agreement decreases. Both schemes can show that the encoder has been modified, while Trans achieves lower agreement. }
\label{ft_agreement}
\end{figure}

\textbf{Fine-Tuning.} 
In the fine-tuning strategy, we considered fine-tuning all levels (FTAL) and compared the agreement obtained by DCmp and Trans during the fine-tuning epochs.

The authentication success rates are \textbf{all 0\%} for encoders pre-trained by ImageNet and CIFAR-10. The agreement is shown in Figure \ref{ft_agreement}. While the number of fine-tuning epochs increases, the changes in the encoder also gradually increase. It can be seen that both DCmp and Trans can reflect the degree to which the encoder has been modified. 

Under the same conditions, Trans yields lower agreement than DCmp, which is in line with our goal of amplifying the small differences in the feature vectors caused by the encoder modification.

\begin{figure}[t]
 \centering
\includegraphics[width=1\linewidth]{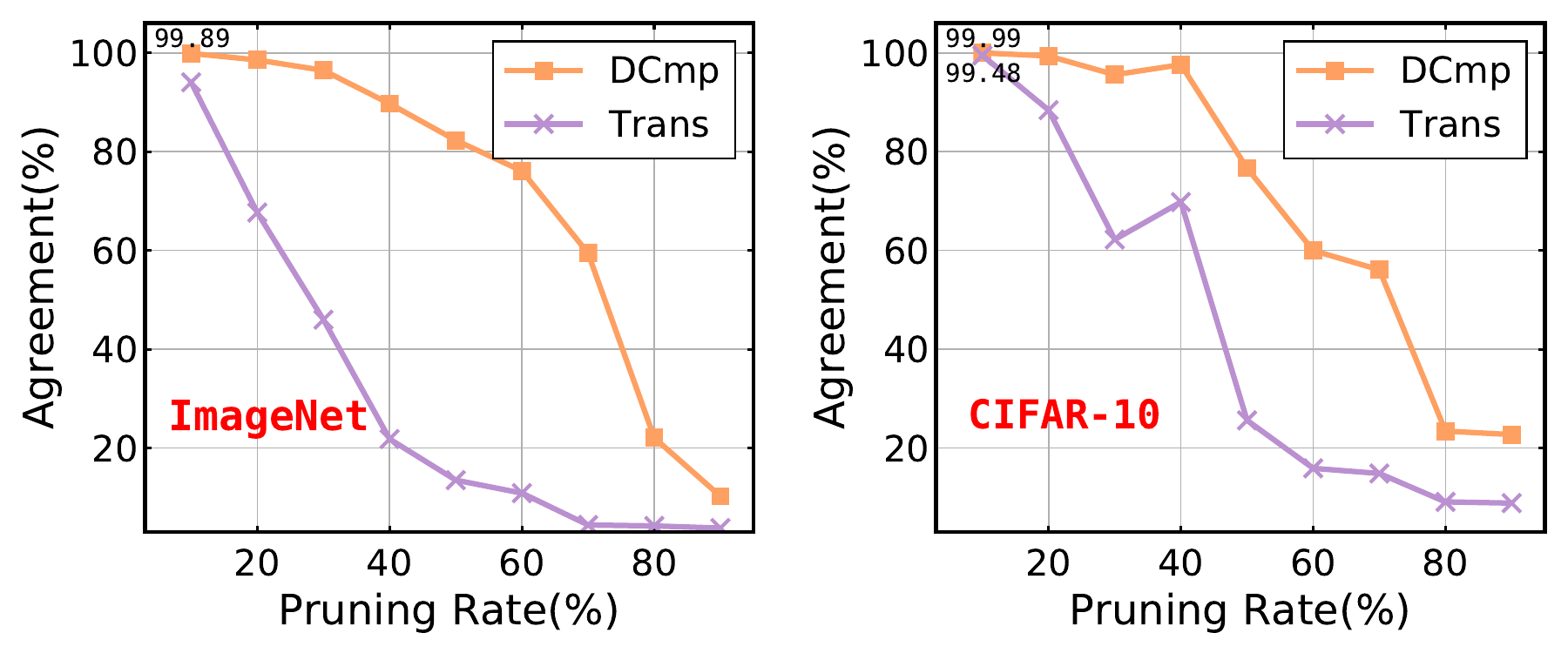}
\caption{Agreement on encoder pruning. The original encoder and pruned encoder with a small pruning rate are very similar, Trans is more suitable for detecting tiny changes of feature vectors. }
\label{pr_agreement}
\end{figure}

\begin{figure}[t]
 \centering
\includegraphics[width=1\linewidth]{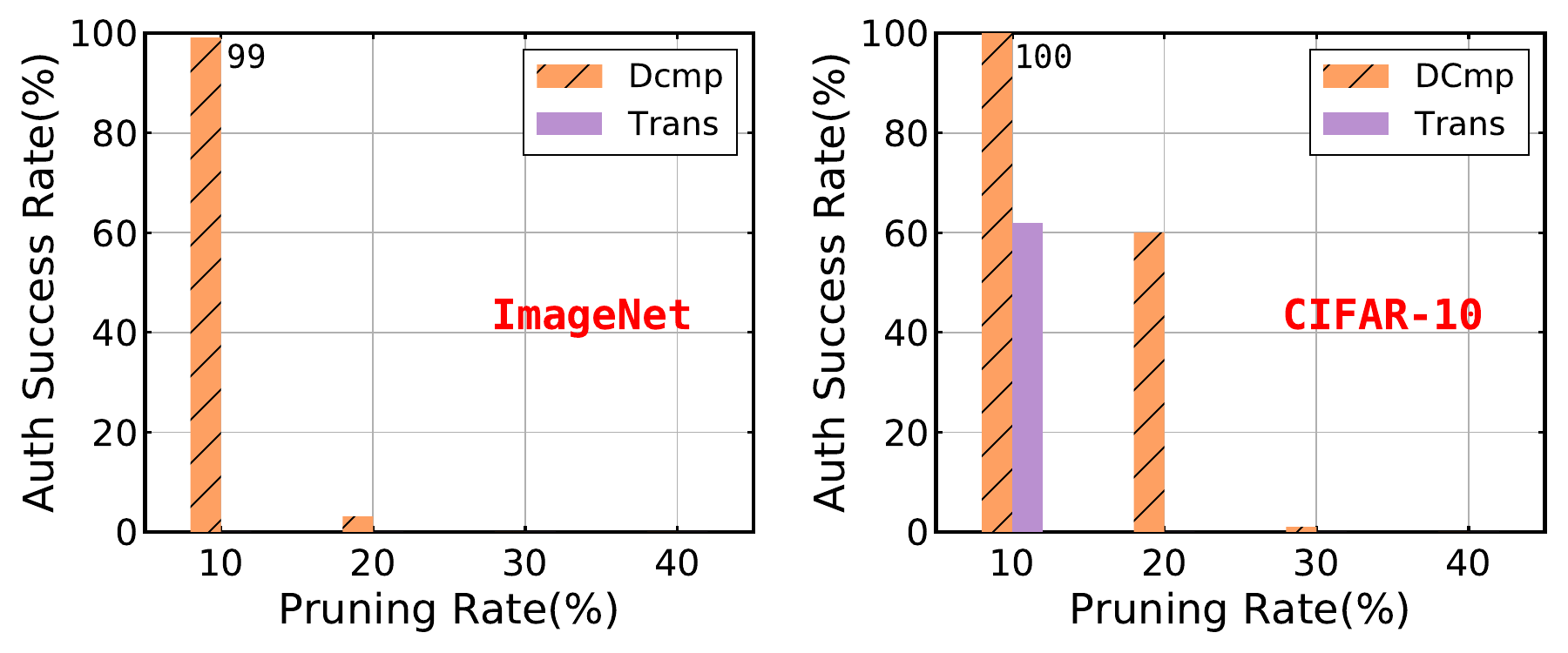}
\caption{Authentication success rate on encoder pruning. When the pruning rate is 10\%, the authentication success rate obtained by DCmp is relatively high, and the similarity of most feature vectors is greater than the threshold constraint $\epsilon$. Trans amplifies the small changes in the feature vectors, achieving better results. }
\label{pr_rate}
\end{figure}

\textbf{Pruning.} 
The pruning process is realized by zeroing out the parameters with the least magnitudes along with a retraining process \cite{han2015deep}, which can significantly reduce the number of parameters in the model without losing too much performance. We sort the parameters of each layer in the encoder based on their absolute values, and set the parameter with a certain proportion of the minimum value in each layer to zero. 

The result of the agreement is shown in Figure \ref{pr_agreement}, and the authentication success rates of different pruning rates are shown in Figure \ref{pr_rate}. For the encoder with pruning rates between 10\% and 90\%, almost all verification failed after pruning. 

When the pruning rate is 10\%, the encoder changes very little. The agreement obtained by DCmp in the ImageNet pre-trained encoder is 99.89\%, and the authentication success rate is 99\%; The agreement obtained from the pre-trained encoder in CIFAR-10 is 99.99\%, and the verification success rate is 100\%. This indicates that DCmp is not sensitive enough to tiny changes in the feature vector caused by changes in the encoder. In this case, Trans has a better performance compared to DCmp. Interestingly, we found that when the pruning rate was 40\%, the agreement corresponding to the CIFAR-10 pre-trained encoder was higher compared to the result with a pruning rate of 30\%. However, when the key samples were selected from a dataset with the same distribution as the pre-trained set, there was no such phenomenon. More results are shown in \ref{A2}.

\begin{table*}[t]
\centering
\renewcommand\arraystretch{1.15}
\begin{tabular}{c|c|cc|cc}
\hline
\multirow{2}{*}{Pre-train Dataset} & \multirow{2}{*}{Target Downstream Dataset} & \multicolumn{2}{c|}{Agreement (\%) $\downarrow$}  & \multicolumn{2}{c}{Auth Success Rate (\%) $\downarrow$} \\ \cline{3-6} 
                          &        & \multicolumn{1}{p{1.7cm}<{\centering}|}{DCmp}  & \multicolumn{1}{p{1.7cm}<{\centering}|}{Trans} & \multicolumn{1}{p{1.7cm}<{\centering}|}{DCmp}  & \multicolumn{1}{p{1.7cm}<{\centering}}{Trans}  \\ \hline
\multirow{3}{*}{ImageNet} & STL-10 & \multicolumn{1}{c|}{23.17}    & 3.23    & \multicolumn{1}{c|}{0}        & 0        \\
                          & GTSRB  & \multicolumn{1}{c|}{22.50}    & 3.21    & \multicolumn{1}{c|}{0}        & 0        \\
                          & SVHN   & \multicolumn{1}{c|}{28.84}    & 3.76     & \multicolumn{1}{c|}{0}        & 0        \\ \hline
\multirow{3}{*}{CIFAR-10}  & STL-10 & \multicolumn{1}{c|}{74.22}    & 21.82    & \multicolumn{1}{c|}{0}        & 0        \\
                          & GTSRB  & \multicolumn{1}{c|}{74.52}    & 21.96    & \multicolumn{1}{c|}{0}        & 0        \\
                          & SVHN   & \multicolumn{1}{c|}{73.87}    & 21.57    & \multicolumn{1}{c|}{0}        & 0        \\ \hline
\end{tabular}
\caption{Agreement and authentication success rate on normal backdoor attacks. According to the knowledge of the adversary, we use CIFAR-10 as the shadow dataset for the CIFAR-10 pre-trained encoder and a 1\% subset of the ImageNet as the shadow dataset for ImageNet pre-trained encoder. Both DCmp and Trans achieve low authentication success rates. }
\label{bd_result}
\end{table*}

\textbf{Backdoor Attack.} 
We followed the backdoor attack method proposed in BadEncoder \cite{jia2022badencoder} and selected several datasets as the target datasets for backdoor attacks. When the downstream dataset was the target dataset, it achieved a high attack success rate ($>$95\%) while ensuring the high accuracy of the downstream classifier. 

We show the agreement and authentication success rate on the backdoor attack in Table \ref{bd_result}. As described in the Section \ref{3}, the adversary $\mathcal{A}$ has the background knowledge of the pre-training dataset that can maximize the similarity between the backdoored encoder and the original encoder. This observation indicates that backdoor attacks have stronger concealment, making these modifications challenging to detect. However, it was not successfully validated in \proposed, revealing the fragility of our fragile watermark. The reason is that the key samples are selected from a dataset with different distributions to the pre-trained dataset, while the adversary can only maintain the performance of the encoder on shadow dataset. 

Based on the above results, \proposed~performs better in fragility on more complex datasets. Protecting such encoders is more necessary as the corresponding encoder construction process requires more resources and effort. 

Similar to the experiment results on fine-tuning and pruning, Trans magnified the differences in feature vectors and obtained lower agreement. Both verification schemes can successfully resist normal backdoor attacks, based on Adversary $\mathcal{A}$'s lack of knowledge about key samples. 

\begin{table}[t]
\centering
\renewcommand\arraystretch{1.15}
\begin{tabular}{c|cc|cc}
\hline
\multirow{2}{*}{\begin{tabular}[c]{@{}c@{}}Pre-train\\  Dataset\end{tabular}} & \multicolumn{2}{c|}{Agreement (\%) $\downarrow$} & \multicolumn{2}{c}{Auth Success Rate (\%) $\downarrow$} \\ \cline{2-5} 
         & \multicolumn{1}{c|}{DCmp}  & Trans & \multicolumn{1}{p{1.4cm}<{\centering}|}{DCmp} & Trans \\ \hline
ImageNet & \multicolumn{1}{c|}{99.98} & 98.20 & \multicolumn{1}{c|}{100}   & 0     \\
CIFAR-10  & \multicolumn{1}{c|}{99.92} & 97.33 & \multicolumn{1}{c|}{100}   & 0     \\ \hline
\end{tabular}
\caption{Agreement and authentication success rate on stronger backdoor attacks with STL-10 as the target downstream task. After obtaining key samples, Adversary $\mathcal{B}$ can disguise the encoder in an attempt to deceive the verification process. This attack can deceive DCmp, but Trans can still distinguish suspicious encoders.}
\label{strong_bd}
\end{table}

\subsection{Performance on Stronger Backdoor Attack}
\label{5.3}
As described in the Section \ref{3} on the adversary's background knowledge, Adversary $\mathcal{B}$ has a very powerful ability. If Adversary $\mathcal{B}$ performs backdoor attacks, it can disguise the backdoor encoder while embedding the backdoor. We added imitation loss to the original loss function of the BadEncoder to show the ability of Adversary $\mathcal{B}$, and the backdoored encoder can learn and imitate the feature vectors of key samples:
\begin{align}
    \mathcal{L}_{imitate} = \mathcal{L}_{cos}(F(\mathbf{K}),F'(\mathbf{K})). 
\end{align}
The final loss function used to execute the backdoor attack is $\mathcal{L}=\mathcal{L}_{original}+\lambda \cdot \mathcal{L}_{imitate}$, here we choose 10 as the value of $\lambda$. More results about it can be seen in \ref{A3}. 

The agreement and authentication success rate on the backdoor attack with imitation loss has been shown in Table \ref{strong_bd}. We set the target downstream task to STL-10 and completed the training of the backdoor encoder with imitation loss. It can be seen that the similarity between the backdoored encoder and the original encoder has increased significantly compared to the result of the normal backdoor attack. This indicates that the imitation operation is effective in deceiving integrity verification. The authentication success rate of DCmp has increased from 0\% to 100\%, but Tran still maintains a verification success rate of 0\%. Obviously, Trans is a better choice to cope with this more deceptive backdoor attack.

\subsection{Comparison on Precision Loss}
\label{5.4}
Furthermore, we have evaluated the effectiveness of our \proposed~on precision loss in Table \ref{compare_eff}. We considered the extreme case of retaining only two decimal places, where DCmp can pass validation, while Trans cannot, as Trans amplifies the small changes in the feature vectors. DCmp can handle the situation of precision loss but will be deceived by Adversary $\mathcal{B}$. In practical applications, we are more likely to choose Trans as the verification scheme because it is less susceptible to deception and can also successfully verify most cases of precision loss.

\begin{table}[t]
\centering
\renewcommand\arraystretch{1.15}
\begin{tabular}{c|cc|cc}
\hline
\multirow{2}{*}{Digit} &
  \multicolumn{2}{c|}{Agreement (\%) $\uparrow$} &
  \multicolumn{2}{c}{Auth Success Rate (\%) $\uparrow$} \\ \cline{2-5} 
                    & \multicolumn{1}{c|}{DCmp}  & Trans  & \multicolumn{1}{p{1.4cm}<{\centering}|}{DCmp} & Trans \\ \hline
 2  & \multicolumn{1}{c|}{99.752} & 94.520  & \multicolumn{1}{c|}{100}  & 0   \\
 3  & \multicolumn{1}{c|}{99.997} & 99.875 & \multicolumn{1}{c|}{100}  & 100   \\ \hline
\end{tabular}
\caption{Effectiveness of \proposed~when precision loss occurs in feature vectors. Almost all cases can pass the verification. }
\label{compare_eff}
\end{table}

\subsection{Comparison to Prior Work}
\label{5.5}

\begin{table}[t]
\centering
\renewcommand\arraystretch{1.15}
\begin{tabular}{c|c|c}
\hline
Method   & Type                    & $\Delta$Accuracy \\
\hline
\cite{cong2022sslguard} & \multirow{5}{*}{Robust} & $\downarrow$ \\
\cite{wu2022watermarking}      &                         & $\downarrow$ \\
\cite{lv2022ssl}      &                         & $\downarrow$ \\
\cite{dziedzic2022dataset}   &                         & = \\
\cite{zhang2023awencoder}   &                         & $\downarrow$ \\
\hline
\proposed     & Fragile                 & =   \\
\hline
\end{tabular}
\caption{Comparison with ``robust'' watermarking methods. Our \proposed~does not cause any performance degradation. }
\label{compare_prior}
\end{table}

To our knowledge, \proposed~is the first work to discuss verifying encoder integrity. Current work on encoder watermarking is aimed at verifying ownership, which is of two types compared to our framework. In addition, as our experiment results indicate, the fragile watermark is very effective in determining whether the encoder has been tampered with, which is different from the robust watermark and can serve as an active defense against malicious modifications. 

Despite these differences, we compared our proposed \proposed~with the previous watermark methods shown in Table \ref{compare_prior}. It can be observed that most of the existing work has had a certain impact on the performance of the encoder, while our proposed \proposed~does not cause any performance degradation.

\section{Discussion}
For the training process of $G$, we have attempted different optimization goals to evaluate the effect on the training process of generative model $G$. We trained three generators represent as $G_{pos}$, $G_{non}$ and $G_{neg}$. $G_{non}$ is an initial generative model without training.
$G_{pos}$ is trained to maximize the similarity between the reconstructed image and the key samples, while $G_{neg}$ is trained to minimize the similarity, both of them are trained for 1000 epochs.

For $G_{pos}$, it can learn the relationship between feature vectors and key samples, similar inputs are likely to be transformed into images similar to key samples, which means that this training target is not sufficient to maximize the distance between outputs of similar inputs.

For $G_{neg}$, compared to $G_{pos}$, minimizing the similarity may make $G$ less dependent on the special mapping with key samples in the feature vector, resulting in greater differences in the output. Making the output not similar to the specific key samples may open up a wider output space, as ``dissimilar'' can have countless manifestations, while ``similar'' typically points to a narrower output range.

For $G_{non}$, an untrained $G$ lacks the ability to map input data to consistent output. Therefore, even if the input data has only slight changes, the similarity between outputs is still random. However, due to the randomness of model parameters, the output generated for any given input (whether similar or different) is random and uncertain. In this case, for similar inputs, the model may produce very different outputs or unexpectedly produce similar outputs, but this is entirely accidental. Due to the fact that the output is mainly randomly generated, untrained GANs are not suitable for applications that require reliable and consistent output.

For the training target of $G$, we hope it can overfit the input feature vectors, and for very similar vectors, it should generate outputs as far as possible. We compared the effects of different training goals in \ref{A1}, and finally chose $G_{neg}$ as the generator for the validation process.

\section{Conclusion}

In this paper, we first introduced the differences between the SL model and SSL encoder in integrity verification, then proposed \proposed, the first encoder integrity verification framework. The framework introduces two validation schemes during the validation phase, directly comparing the feature vectors of the key samples or converting them into image domain. The first scheme is more convenient to implement, while the second scheme is more capable of capturing tiny changes in feature vectors of key samples. Both schemes can effectively detect whether the encoder has been modified.

\clearpage
\bibliographystyle{aaai}
\bibliography{main}

\begin{thebibliography}{}

\bibitem[\protect\citeauthoryear{Adi \bgroup et al\mbox.\egroup
  }{2018}]{adi2018turning}
Adi, Y.; Baum, C.; Cisse, M.; Pinkas, B.; and Keshet, J.
\newblock 2018.
\newblock Turning your weakness into a strength: Watermarking deep neural
  networks by backdooring.
\newblock In {\em 27th USENIX Security Symposium (USENIX Security)},
  1615--1631.

\bibitem[\protect\citeauthoryear{Botta, Cavagnino, and
  Esposito}{2021}]{botta2021neunac}
Botta, M.; Cavagnino, D.; and Esposito, R.
\newblock 2021.
\newblock Neunac: A novel fragile watermarking algorithm for integrity
  protection of neural networks.
\newblock {\em Information Sciences} 576:228--241.

\bibitem[\protect\citeauthoryear{Carlini and
  Terzis}{2021}]{carlini2021poisoning}
Carlini, N., and Terzis, A.
\newblock 2021.
\newblock Poisoning and backdooring contrastive learning.
\newblock {\em arXiv preprint arXiv:2106.09667}.

\bibitem[\protect\citeauthoryear{Chen \bgroup et al\mbox.\egroup
  }{2019}]{chen2019deepmarks}
Chen, H.; Rouhani, B.~D.; Fu, C.; Zhao, J.; and Koushanfar, F.
\newblock 2019.
\newblock Deepmarks: A secure fingerprinting framework for digital rights
  management of deep learning models.
\newblock In {\em Proceedings of the 2019 on International Conference on
  Multimedia Retrieval},  105--113.

\bibitem[\protect\citeauthoryear{Chen \bgroup et al\mbox.\egroup
  }{2020a}]{chen2020simple}
Chen, T.; Kornblith, S.; Norouzi, M.; and Hinton, G.
\newblock 2020a.
\newblock A simple framework for contrastive learning of visual
  representations.
\newblock In {\em International Conference on Machine Learning},  1597--1607.
\newblock PMLR.

\bibitem[\protect\citeauthoryear{Chen \bgroup et al\mbox.\egroup
  }{2020b}]{chen2020improved}
Chen, X.; Fan, H.; Girshick, R.; and He, K.
\newblock 2020b.
\newblock Improved baselines with momentum contrastive learning.
\newblock {\em arXiv preprint arXiv:2003.04297}.

\bibitem[\protect\citeauthoryear{Coates, Ng, and
  Lee}{2011}]{coates2011analysis}
Coates, A.; Ng, A.; and Lee, H.
\newblock 2011.
\newblock An analysis of single-layer networks in unsupervised feature
  learning.
\newblock In {\em Proceedings of the fourteenth international conference on
  artificial intelligence and statistics},  215--223.
\newblock JMLR Workshop and Conference Proceedings.

\bibitem[\protect\citeauthoryear{Cong, He, and Zhang}{2022}]{cong2022sslguard}
Cong, T.; He, X.; and Zhang, Y.
\newblock 2022.
\newblock Sslguard: A watermarking scheme for self-supervised learning
  pre-trained encoders.
\newblock In {\em Proceedings of the 2022 ACM SIGSAC Conference on Computer and
  Communications Security},  579--593.

\bibitem[\protect\citeauthoryear{Darvish~Rouhani, Chen, and
  Koushanfar}{2019}]{darvish2019deepsigns}
Darvish~Rouhani, B.; Chen, H.; and Koushanfar, F.
\newblock 2019.
\newblock Deepsigns: An end-to-end watermarking framework for ownership
  protection of deep neural networks.
\newblock In {\em Proceedings of the Twenty-Fourth International Conference on
  Architectural Support for Programming Languages and Operating Systems},
  485--497.

\bibitem[\protect\citeauthoryear{Devlin \bgroup et al\mbox.\egroup
  }{2018}]{devlin2018bert}
Devlin, J.; Chang, M.-W.; Lee, K.; and Toutanova, K.
\newblock 2018.
\newblock Bert: Pre-training of deep bidirectional transformers for language
  understanding.
\newblock {\em arXiv preprint arXiv:1810.04805}.

\bibitem[\protect\citeauthoryear{Dziedzic \bgroup et al\mbox.\egroup
  }{2022a}]{dziedzic2022difficulty}
Dziedzic, A.; Dhawan, N.; Kaleem, M.~A.; Guan, J.; and Papernot, N.
\newblock 2022a.
\newblock On the difficulty of defending self-supervised learning against model
  extraction.
\newblock In {\em International Conference on Machine Learning},  5757--5776.
\newblock PMLR.

\bibitem[\protect\citeauthoryear{Dziedzic \bgroup et al\mbox.\egroup
  }{2022b}]{dziedzic2022dataset}
Dziedzic, A.; Duan, H.; Kaleem, M.~A.; Dhawan, N.; Guan, J.; Cattan, Y.;
  Boenisch, F.; and Papernot, N.
\newblock 2022b.
\newblock Dataset inference for self-supervised models.
\newblock {\em Proceedings of Advances in Neural Information Processing
  Systems} 35:12058--12070.

\bibitem[\protect\citeauthoryear{Goodfellow \bgroup et al\mbox.\egroup
  }{2014}]{goodfellow2014generative}
Goodfellow, I.; Pouget-Abadie, J.; Mirza, M.; Xu, B.; Warde-Farley, D.; Ozair,
  S.; Courville, A.; and Bengio, Y.
\newblock 2014.
\newblock Generative adversarial nets.
\newblock In {\em Proceedings of Advances in Neural Information Processing
  Systems}.

\bibitem[\protect\citeauthoryear{Guan \bgroup et al\mbox.\egroup
  }{2020}]{guan2020reversible}
Guan, X.; Feng, H.; Zhang, W.; Zhou, H.; Zhang, J.; and Yu, N.
\newblock 2020.
\newblock Reversible watermarking in deep convolutional neural networks for
  integrity authentication.
\newblock In {\em Proceedings of the 28th ACM International Conference on
  Multimedia},  2273--2280.

\bibitem[\protect\citeauthoryear{Han, Mao, and Dally}{2015}]{han2015deep}
Han, S.; Mao, H.; and Dally, W.~J.
\newblock 2015.
\newblock Deep compression: Compressing deep neural networks with pruning,
  trained quantization and huffman coding.
\newblock {\em arXiv preprint arXiv:1510.00149}.

\bibitem[\protect\citeauthoryear{He \bgroup et al\mbox.\egroup
  }{2020}]{he2020momentum}
He, K.; Fan, H.; Wu, Y.; Xie, S.; and Girshick, R.
\newblock 2020.
\newblock Momentum contrast for unsupervised visual representation learning.
\newblock In {\em Proceedings of the IEEE/CVF Conference on Computer Vision and
  Pattern Recognition},  9729--9738.

\bibitem[\protect\citeauthoryear{Hjelm \bgroup et al\mbox.\egroup
  }{2018}]{hjelm2018learning}
Hjelm, R.~D.; Fedorov, A.; Lavoie-Marchildon, S.; Grewal, K.; Bachman, P.;
  Trischler, A.; and Bengio, Y.
\newblock 2018.
\newblock Learning deep representations by mutual information estimation and
  maximization.
\newblock {\em arXiv preprint arXiv:1808.06670}.

\bibitem[\protect\citeauthoryear{Jia, Liu, and Gong}{2022}]{jia2022badencoder}
Jia, J.; Liu, Y.; and Gong, N.~Z.
\newblock 2022.
\newblock Badencoder: Backdoor attacks to pre-trained encoders in
  self-supervised learning.
\newblock In {\em 2022 IEEE Symposium on Security and Privacy (SP)},
  2043--2059.
\newblock IEEE.

\bibitem[\protect\citeauthoryear{Karras \bgroup et al\mbox.\egroup
  }{2017}]{karras2017progressive}
Karras, T.; Aila, T.; Laine, S.; and Lehtinen, J.
\newblock 2017.
\newblock Progressive growing of gans for improved quality, stability, and
  variation.
\newblock {\em arXiv preprint arXiv:1710.10196}.

\bibitem[\protect\citeauthoryear{Krizhevsky and
  Hinton}{2009}]{krizhevsky2009learning}
Krizhevsky, A., and Hinton, G.
\newblock 2009.
\newblock Learning multiple layers of features from tiny images.
\newblock {\em Tech Report}.

\bibitem[\protect\citeauthoryear{Lao \bgroup et al\mbox.\egroup
  }{2022}]{lao2022deepauth}
Lao, Y.; Zhao, W.; Yang, P.; and Li, P.
\newblock 2022.
\newblock Deepauth: A dnn authentication framework by model-unique and fragile
  signature embedding.
\newblock In {\em Proceedings of the AAAI Conference on Artificial
  Intelligence}, volume~36,  9595--9603.

\bibitem[\protect\citeauthoryear{Li \bgroup et al\mbox.\egroup
  }{2022}]{li2022demystifying}
Li, C.; Pang, R.; Xi, Z.; Du, T.; Ji, S.; Yao, Y.; and Wang, T.
\newblock 2022.
\newblock Demystifying self-supervised trojan attacks.
\newblock {\em arXiv preprint arXiv:2210.07346}.

\bibitem[\protect\citeauthoryear{Liu \bgroup et al\mbox.\egroup
  }{2022}]{liu2022stolenencoder}
Liu, Y.; Jia, J.; Liu, H.; and Gong, N.~Z.
\newblock 2022.
\newblock Stolenencoder: stealing pre-trained encoders in self-supervised
  learning.
\newblock In {\em Proceedings of the 2022 ACM SIGSAC Conference on Computer and
  Communications Security},  2115--2128.

\bibitem[\protect\citeauthoryear{Lv \bgroup et al\mbox.\egroup
  }{2022}]{lv2022ssl}
Lv, P.; Li, P.; Zhu, S.; Zhang, S.; Chen, K.; Liang, R.; Yue, C.; Xiang, F.;
  Cai, Y.; Ma, H.; et~al.
\newblock 2022.
\newblock Ssl-wm: A black-box watermarking approach for encoders pre-trained by
  self-supervised learning.
\newblock {\em arXiv preprint arXiv:2209.03563}.

\bibitem[\protect\citeauthoryear{Netzer \bgroup et al\mbox.\egroup
  }{2011}]{netzer2011reading}
Netzer, Y.; Wang, T.; Coates, A.; Bissacco, A.; Wu, B.; and Ng, A.~Y.
\newblock 2011.
\newblock Reading digits in natural images with unsupervised feature learning.
\newblock In {\em Proceedings of the NIPS Workshop on Deep Learning and
  Unsupervised Feature Learning}.

\bibitem[\protect\citeauthoryear{Peng \bgroup et al\mbox.\egroup
  }{2022}]{peng2022fingerprinting}
Peng, Z.; Li, S.; Chen, G.; Zhang, C.; Zhu, H.; and Xue, M.
\newblock 2022.
\newblock Fingerprinting deep neural networks globally via universal
  adversarial perturbations.
\newblock In {\em Proceedings of the IEEE/CVF Conference on Computer Vision and
  Pattern Recognition},  13430--13439.

\bibitem[\protect\citeauthoryear{Russakovsky \bgroup et al\mbox.\egroup
  }{2015}]{russakovsky2015imagenet}
Russakovsky, O.; Deng, J.; Su, H.; Krause, J.; Satheesh, S.; Ma, S.; Huang, Z.;
  Karpathy, A.; Khosla, A.; Bernstein, M.; et~al.
\newblock 2015.
\newblock Imagenet large scale visual recognition challenge.
\newblock {\em International Journal of Computer Vision} 115:211--252.

\bibitem[\protect\citeauthoryear{Saha \bgroup et al\mbox.\egroup
  }{2022}]{saha2022backdoor}
Saha, A.; Tejankar, A.; Koohpayegani, S.~A.; and Pirsiavash, H.
\newblock 2022.
\newblock Backdoor attacks on self-supervised learning.
\newblock In {\em Proceedings of the IEEE/CVF Conference on Computer Vision and
  Pattern Recognition},  13337--13346.

\bibitem[\protect\citeauthoryear{Sha \bgroup et al\mbox.\egroup
  }{2023}]{sha2023can}
Sha, Z.; He, X.; Yu, N.; Backes, M.; and Zhang, Y.
\newblock 2023.
\newblock Can't steal? cont-steal! contrastive stealing attacks against image
  encoders.
\newblock In {\em Proceedings of the IEEE/CVF Conference on Computer Vision and
  Pattern Recognition},  16373--16383.

\bibitem[\protect\citeauthoryear{Stallkamp \bgroup et al\mbox.\egroup
  }{2012}]{stallkamp2012man}
Stallkamp, J.; Schlipsing, M.; Salmen, J.; and Igel, C.
\newblock 2012.
\newblock Man vs. computer: Benchmarking machine learning algorithms for
  traffic sign recognition.
\newblock {\em Neural networks} 32:323--332.

\bibitem[\protect\citeauthoryear{Uchida \bgroup et al\mbox.\egroup
  }{2017}]{uchida2017embedding}
Uchida, Y.; Nagai, Y.; Sakazawa, S.; and Satoh, S.
\newblock 2017.
\newblock Embedding watermarks into deep neural networks.
\newblock In {\em Proceedings of the 2017 ACM on International Conference on
  Multimedia Retrieval},  269--277.

\bibitem[\protect\citeauthoryear{Wang \bgroup et al\mbox.\egroup
  }{2004}]{wang2004image}
Wang, Z.; Bovik, A.~C.; Sheikh, H.~R.; and Simoncelli, E.~P.
\newblock 2004.
\newblock Image quality assessment: from error visibility to structural
  similarity.
\newblock {\em IEEE Transactions on Image Processing} 13(4):600--612.

\bibitem[\protect\citeauthoryear{Wu \bgroup et al\mbox.\egroup
  }{2022}]{wu2022watermarking}
Wu, Y.; Qiu, H.; Zhang, T.; Li, J.; and Qiu, M.
\newblock 2022.
\newblock Watermarking pre-trained encoders in contrastive learning.
\newblock In {\em 2022 4th International Conference on Data Intelligence and
  Security (ICDIS)},  228--233.
\newblock IEEE.

\bibitem[\protect\citeauthoryear{Xue and Lou}{2022}]{xue2022estas}
Xue, J., and Lou, Q.
\newblock 2022.
\newblock Estas: Effective and stable trojan attacks in self-supervised
  encoders with one target unlabelled sample.
\newblock {\em arXiv preprint arXiv:2211.10908}.

\bibitem[\protect\citeauthoryear{Yin, Yin, and Zhang}{2022}]{yin2022neural}
Yin, Z.; Yin, H.; and Zhang, X.
\newblock 2022.
\newblock Neural network fragile watermarking with no model performance
  degradation.
\newblock In {\em 2022 IEEE International Conference on Image Processing
  (ICIP)},  3958--3962.
\newblock IEEE.

\bibitem[\protect\citeauthoryear{Zhang \bgroup et al\mbox.\egroup
  }{2023}]{zhang2023awencoder}
Zhang, T.; Wu, H.; Lu, X.; Han, G.; and Sun, G.
\newblock 2023.
\newblock Awencoder: Adversarial watermarking pre-trained encoders in
  contrastive learning.
\newblock {\em Applied Sciences} 13(6):3531.

\bibitem[\protect\citeauthoryear{Zhong \bgroup et al\mbox.\egroup
  }{2020}]{zhong2020protecting}
Zhong, Q.; Zhang, L.~Y.; Zhang, J.; Gao, L.; and Xiang, Y.
\newblock 2020.
\newblock Protecting ip of deep neural networks with watermarking: A new label
  helps.
\newblock In {\em Advances in Knowledge Discovery and Data Mining: 24th
  Pacific-Asia Conference, PAKDD 2020, Singapore, May 11--14, 2020,
  Proceedings, Part II 24},  462--474.
\newblock Springer.

\bibitem[\protect\citeauthoryear{Zhu \bgroup et al\mbox.\egroup
  }{2021}]{zhu2021fragile}
Zhu, R.; Wei, P.; Li, S.; Yin, Z.; Zhang, X.; and Qian, Z.
\newblock 2021.
\newblock Fragile neural network watermarking with trigger image set.
\newblock In {\em Knowledge Science, Engineering and Management: 14th
  International Conference, KSEM 2021, Tokyo, Japan, August 14--16, 2021,
  Proceedings, Part I 14},  280--293.
\newblock Springer.

\end{thebibliography}

\appendix
\clearpage

\section{Supplementary Material}
In this supplementary material, we present a comprehensive understanding and analysis of our proposed method, along with additional experimental results for a thorough evaluation.

\begin{itemize}
\item We examine and analyze the impact of the generator's training objectives on integrity verification.

\item A series of experiments are conducted on the selection of key samples, investigating the influence of the distribution of these key samples on the proposed~\proposed.

\item A detailed explanation is provided for the process through which the adversary camouflaged the backdoor encoder. Additionally, we compare the effectiveness of the verification scheme on the disguised backdoor encoder.

\item We discuss and compare various implementation details of the verification scheme, demonstrating their efficacy in integrity verification.
\end{itemize}

\subsection{Different Training Process of Generator}
\label{A1}
We evaluate the performance of the generator on a backdoor encoder targeting STL-10 downstream tasks, and $G_{neg}$ achieves the best overall results for the carefully disguised backdoor encoder.

The impact of the training process on the performance of the generator is shown in Table \ref{train_imitate}, with the untrained $G_{non}$ exhibiting randomness in its results. $G_{non}$ performs optimally on the CIFAR-10 pre-trained backdoor encoder while simultaneously achieving the highest agreement on the ImageNet pre-trained backdoor encoder which is not ideal.

As the training goal of $G_{pos}$ is to transform the feature vectors of key samples into images $G_{pos}(F(\mathbf{K}))$ similar to the key samples $\mathbf{K}$, similar inputs may result in images with analogous characteristics. In comparison to $G_{neg}$, $G_{pos}$ produces more similar outputs for similar inputs, leading to a higher agreement and authentication success rate on the disguised encoder. We observe that $G_{neg}$ maximizes the small differences in feature vectors, consistently maintaining stable performance with a low authentication success rate.

\begin{table}[]
\centering
\renewcommand\arraystretch{1.15}
\begin{tabular}{c|c|p{1.9cm}<{\centering}|p{1.7cm}<{\centering}}
\hline
\begin{tabular}[c]{@{}c@{}}Pre-train\\ Dataset\end{tabular} & Scheme & \begin{tabular}[c]{@{}c@{}}Agreement \\(\%) $\downarrow$\end{tabular} & \begin{tabular}[c]{@{}c@{}}Auth Success \\ Rate (\%) $\downarrow$\end{tabular} \\ \hline
\multirow{4}{*}{ImageNet} & DCmp       & 99.98 & 100 \\
                          & $G_{pos}$ & 98.95 & 14   \\
                          & $G_{non}$ & 99.20 & 18  \\
                          & $G_{neg}$ & \textbf{98.2}  & \textbf{0}   \\ \hline
\multirow{4}{*}{CIFAR-10} & DCmp       & 99.92 & 100 \\
                          & $G_{pos}$ & 99.07 & 16  \\
                          & $G_{non}$ & \textbf{97.83} & \textbf{0}   \\
                          & $G_{neg}$ & 97.98 & \textbf{0}   \\ \hline
\end{tabular}
\caption{Comparison on disguised backdoor encoder with STL-10 as the target downstream task. $G_{non}$ better detect changes in the CIFAR-10 pre-trained encoder, while performing poorly on the ImageNet pre-trained encoder, demonstrating its randomness. Although $G_{pos}$ can also reflect changes in the encoder, $G_{neg}$ ensures a lower verification success rate.}
\label{train_imitate}
\end{table}

\subsection{Key Samples in Different Distributions}
\label{A2}

To assess the influence of key samples on the verification process, we compare schemes for selecting key samples and subsequently evaluate them with different distributions. Key samples sharing the same distribution as the pre-training dataset are termed In-Distribution (ID), while those with distributions differing from the pre-training dataset are referred to as Out-Of-Distribution (OOD). We first explore the selection of key samples from ID data.

\begin{figure}[]
 \centering
\includegraphics[width=1\linewidth]{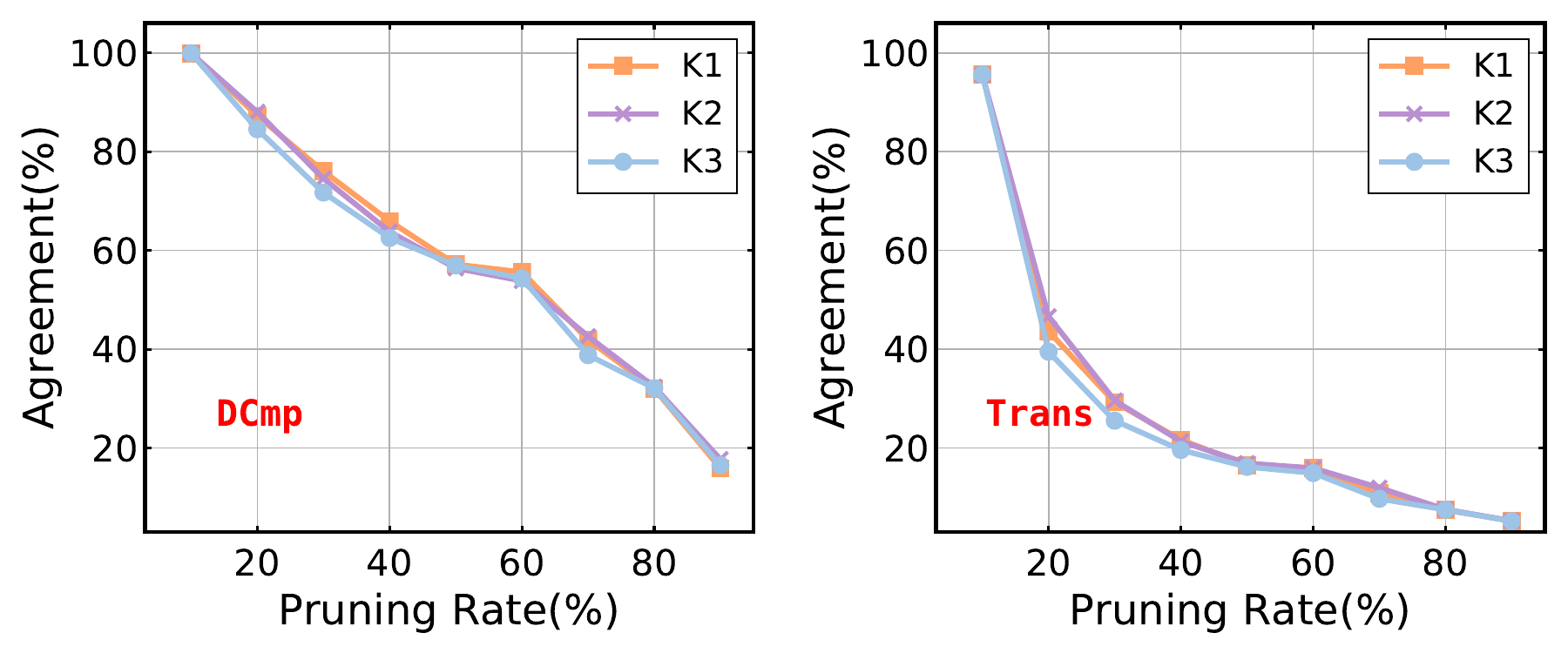}
\caption{Different selections of ID key samples achieve similar performance when the pruning encoder. DCmp and Trans also obtain similar agreement.}
\label{k123}
\end{figure}

Given that the encoder doesn't utilize category labels during the pre-training process, we randomly choose 5 samples from each class within the 20-class subset of ImageNet as key samples $K_1$. Subsequently, we select 100 samples from a randomly chosen class, denoted as $K_2$. Additionally, one sample is randomly chosen from each of the 100 randomly sampled classes as $K_3$. The effectiveness of different key samples on the ImageNet pre-trained encoder is evaluated in Figure \ref{k123}.

As observed, results obtained from ID key samples generally exhibit high similarity. Consequently, we assert that, within the same distribution, the strategy employed for selecting key samples is unlikely to exert a significant impact on the verification process.

\begin{figure}[t]
 \centering
\includegraphics[width=1\linewidth]{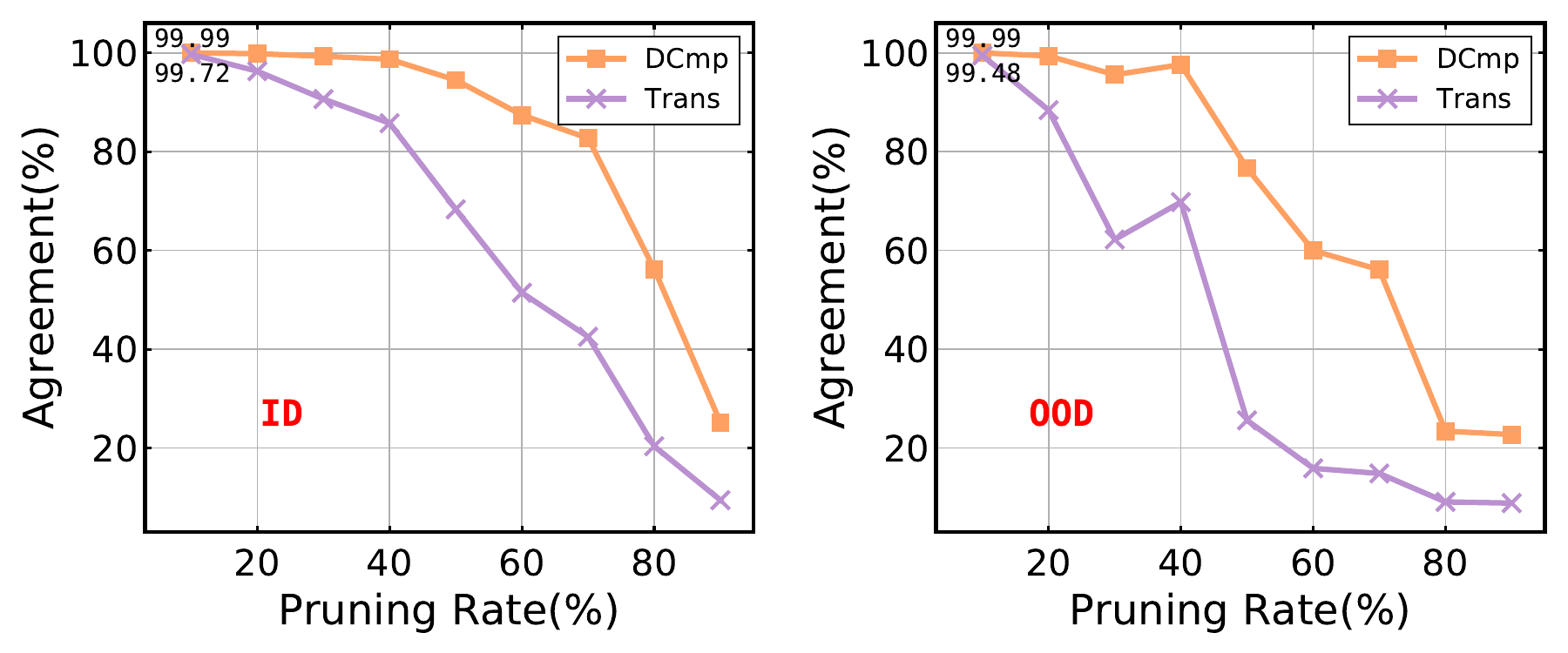}
\caption{Results of pruning CIFAR-10 pre-trained encoder. Key samples with the same distribution as the pre-training dataset exhibit no significant changes, while OOD samples obtain less agreement when the encoder is modified.}
\label{dis_agreement}
\end{figure}

\begin{figure}[t]
 \centering
\includegraphics[width=1\linewidth]{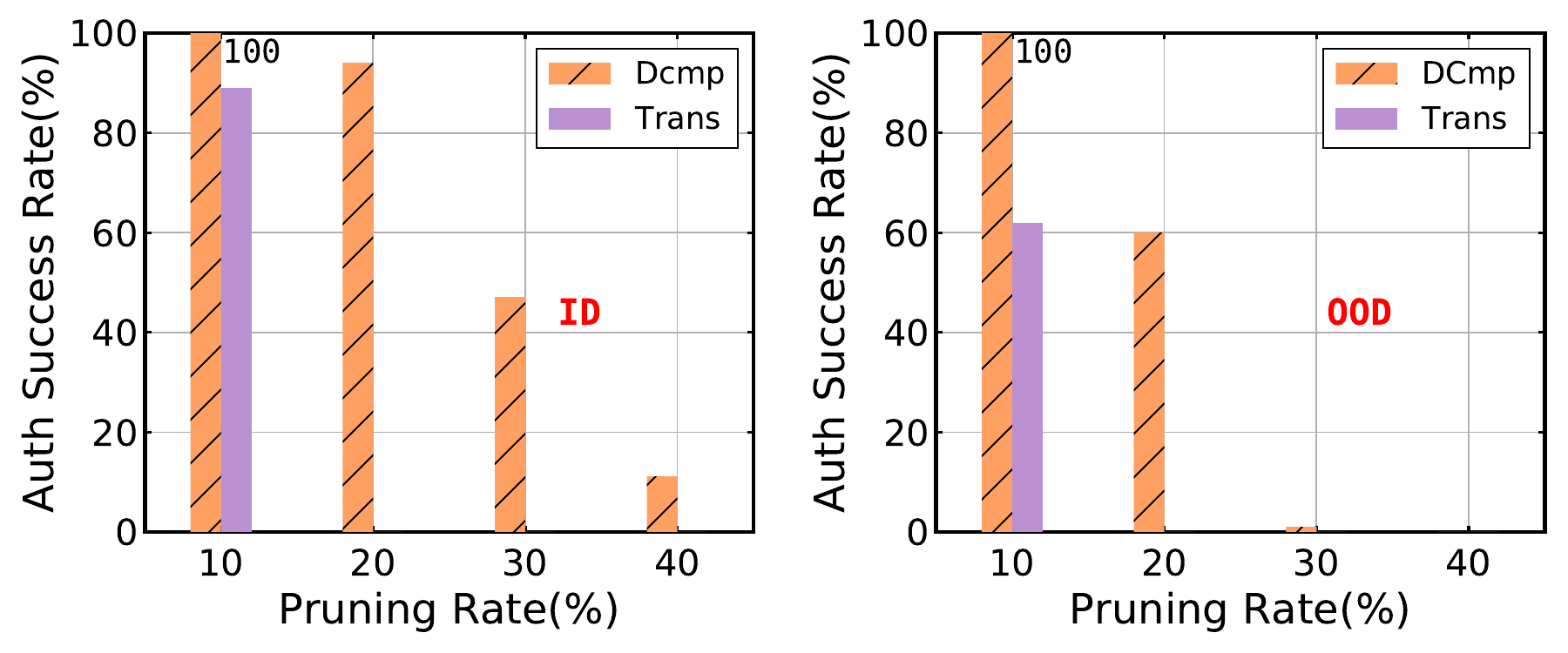}
\caption{Comparing the authentication success rate of key samples in different distributions. OOD samples obtain lower authentication success rates on encoder pruning, and both DCmp and Trans benefit from it.}
\label{dis_rate}
\end{figure}

Furthermore, we conduct a comparison involving randomly selected key samples from OOD data. Specifically, for the CIFAR-10 pre-trained encoder, we choose key samples from ImageNet. The outcomes, depicted in Figure \ref{dis_agreement} and Figure \ref{dis_rate}, illustrate the impact of key samples on the verification process when applying encoder pruning to the CIFAR-10 pre-trained encoder.

\begin{figure*}[t]
 \centering
\includegraphics[width=1\linewidth]{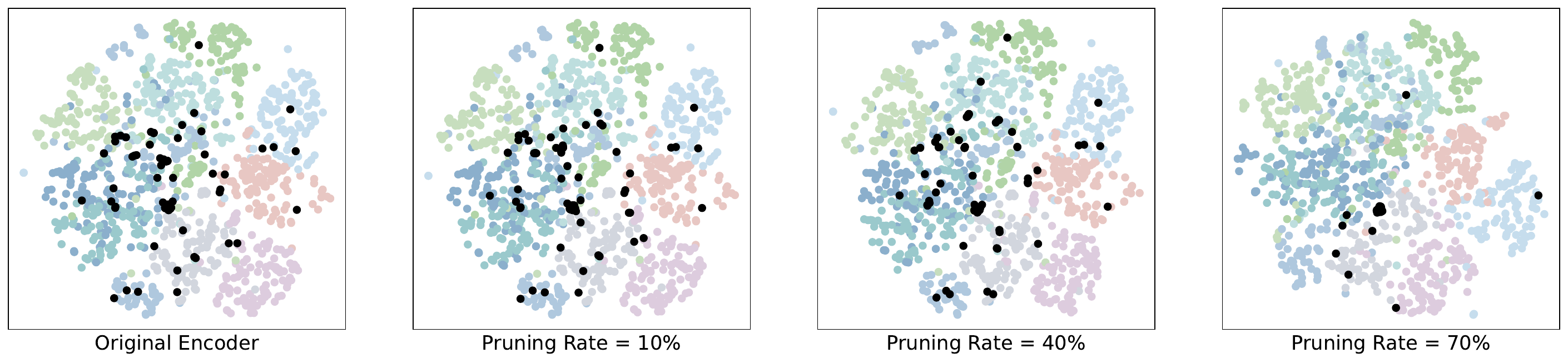}
\caption{Visualization of feature domain. With the CIFAR-10 pre-trained encoder, we evaluate the change in features. The feature of ID data maintains a very similar distribution to the original. Meanwhile, the changes in features obtained by OOD key samples (points in black) are more chaotic, with more offsets occurring compared to ID data.}
\label{dis}
\end{figure*}

When key samples share the same distribution as the pre-training dataset, the encoder, having learned from this distribution, tends to maintain relatively similar outputs even after modification. However, selecting key samples with a distribution different from that of the pre-training dataset leads to increased uncertainty in the outputs when the encoder is modified. We depict the changes in distributions after pruning on CIFAR-10 pre-trained encoders in Figure \ref{dis}.

For the case of precision loss, the performance of ID data is similar to that of OOD data. The alterations in feature vectors caused by precision loss exhibit similar patterns across different sample distributions. As illustrated in Table \ref{precision}, it is evident that precision loss in key samples with different distributions does not impact the verification process. Consequently, OOD data proves to be more effective in revealing malicious tampering of the encoder without influencing the verification results related to precision loss. Opting for key samples with a distribution different from that of the pre-training dataset proves advantageous for verification purposes.

\begin{table}[t]
\centering
\renewcommand\arraystretch{1.15}
\begin{tabular}{c|c|cc|cc}
\hline
\multirow{2}{*}{$D_K$} &
  \multirow{2}{*}{Digit} &
  \multicolumn{2}{c|}{Agreement (\%) $\uparrow$} &
  \multicolumn{2}{c}{\begin{tabular}[c]{@{}c@{}}Auth Success \\ Rate (\%) $\uparrow$ \end{tabular}} \\ \cline{3-6} 
                          &   & \multicolumn{1}{c|}{DCmp}   & Trans & \multicolumn{1}{c|}{DCmp}   & Trans \\ \hline
\multirow{2}{*}{ImageNet} & 2 & \multicolumn{1}{c|}{99.752} & 94.520   & \multicolumn{1}{c|}{100} & 0     \\
                          & 3 & \multicolumn{1}{c|}{99.997} & 99.875   & \multicolumn{1}{c|}{100} & 100   \\ \hline
\multirow{2}{*}{CIFAR-10} & 2 & \multicolumn{1}{c|}{99.762} & 94.993   & \multicolumn{1}{c|}{100} & 0     \\
                          & 3 & \multicolumn{1}{c|}{99.997} & 99.885   & \multicolumn{1}{c|}{100} & 100   \\ \hline
\end{tabular}
\caption{Comparing the results of precision loss on key samples in different distributions. We choose the CIFAR-10 pre-trained encoder to show the results, OOD samples (ImageNet as $D_K$) perform likely to ID samples (CIFAR-10 as $D_K$).}
\label{precision}
\end{table}

\subsection{Results on Disguised Backdoor Encoders}
\label{A3}
To showcase the capabilities of adversary $\mathcal{B}$, we introduce an imitation loss into the backdoor attack process in Section \ref{5.3}, aiming at preserving the original performance of key samples and deceiving the integrity verification of the encoder. A parameter $\lambda$ is employed to regulate the weight of the imitation loss during the backdoor embedding.

\begin{figure}[t]
 \centering
\includegraphics[width=0.7\linewidth]{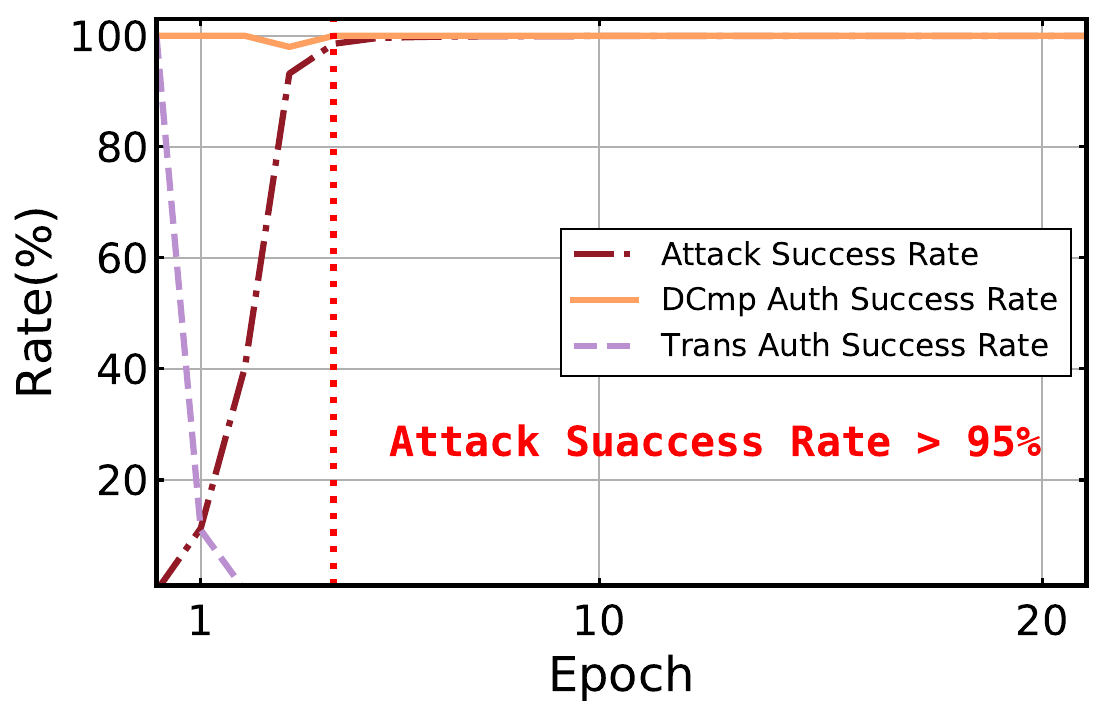}
\caption{Authentication success rate and attack success rate in the backdoor embedding process. Adversary $\mathcal{B}$ can imitate the original outputs of key samples, and DCmp would be deceived by the disguised backdoor encoder, while Trans can detect this kind of backdoor attack easily.}
\label{asr_rate}
\end{figure}

Initially, we set $\lambda=10$ and embed backdoors on the clean encoder pre-trained on CIFAR-10, targeting STL-10 as the downstream task. In Figure \ref{asr_rate}, the outcomes illustrate the variations in authentication success rate and attack success rate during the execution of the backdoor attack with imitation loss. 
Adversary $\mathcal{B}$ can minimize the distance between the outputs of the backdoor encoder for the key samples and the original outputs, which enables training a backdoor encoder with a high attack success rate that can bypass key sample verification. 

Despite adversary $\mathcal{B}$ successfully obtaining a backdoor encoder that meets the specified goal through imitation loss, and DCmp being deceived, Trans achieves a low authentication success rate of 11\% when executing only one epoch of backdoor embedding. This enables Trans to successfully identify the well-disguised backdoor encoder.

\begin{figure}[t]
 \centering
\includegraphics[width=0.7\linewidth]{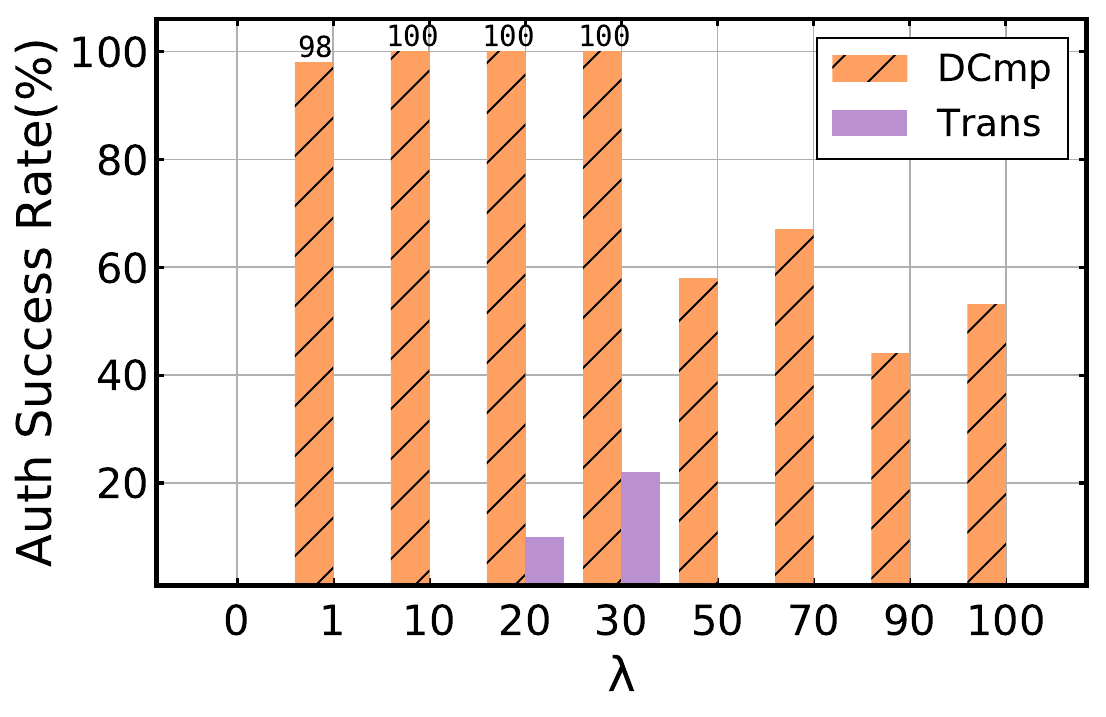}
\caption{The impact of utilizing different $\lambda$ when embedding backdoor. A large value of $\lambda$ does not lead to a perfectly disguised backdoor encoder. Trans always maintains a low authentication success rate on well-disguised backdoor encoders.}
\label{lambda}
\end{figure}

To explore the impact of the hyperparameter $\lambda$ on the verification process, we conduct a series of experiments using various values of $\lambda$ on the ImageNet pre-trained backdoor encoder with STL-10 as the target downstream task. As depicted in Figure \ref{lambda}, it becomes apparent that the authentication success rate does not consistently uphold a high level with the escalation of $\lambda$. Excessive values of $\lambda$ will affect the backdoor embedding process.
Even when the backdoor encoder attains the pinnacle of imitation, Trans effectively thwarts the disguised backdoor encoder from circumventing the verification process. In scenarios involving adversaries possessing significant background knowledge, the adoption of Trans as a robust integrity verification scheme is essential for fortifying security and upholding users' rights.

Furthermore, we assess the efficacy of the imitation loss across diverse downstream tasks. As portrayed in Table \ref{all_imitate}, imitation loss demonstrates its prowess in concealing backdoor encoders, showcasing adaptability across a spectrum of target downstream tasks. Regrettably, DCmp falters in distinguishing the disguised backdoor encoder, whereas Trans consistently maintains a formidable performance. In practical applications, for mitigating the potential threat of backdoor attacks, Trans emerges as the more fitting choice for an integrity verification scheme.

\begin{table*}[t]
\centering
\renewcommand\arraystretch{1.15}
\begin{tabular}{c|c|cc|cc}
\hline
\multirow{2}{*}{Target Downstream Task} &
  \multirow{2}{*}{Attack Success Rate (\%)} &
  \multicolumn{2}{p{4cm}<{\centering}|}{Agreement (\%) $\downarrow$} &
  \multicolumn{2}{p{4cm}<{\centering}}{Auth Success Rate (\%) $\downarrow$} \\ \cline{3-6} 
      &       & \multicolumn{1}{p{2cm}<{\centering}|}{DCmp}  & Trans & \multicolumn{1}{p{2cm}<{\centering}|}{DCmp} & Trans \\ \hline
STL10 & 99.56 & \multicolumn{1}{c|}{99.92} & 97.33 & \multicolumn{1}{c|}{100}  & 0     \\
GTSRB & 95.33 & \multicolumn{1}{c|}{99.92} & 97.21 & \multicolumn{1}{c|}{100}  & 0     \\
SVHN  & 97.82 & \multicolumn{1}{c|}{99.91} & 96.73 & \multicolumn{1}{c|}{100}  & 0     \\ \hline
\end{tabular}
\caption{Disguised CIFAR-10 pre-trained encoder with different target downstream tasks. The imitation loss can effectively deceive DCmp in various downstream tasks. Trans can always achieve stable performance with a low authentication success rate.}
\label{all_imitate}
\end{table*}

\subsection{Verification Schemes Comparison}
\label{A4}

\begin{table*}[t]
\centering
\renewcommand\arraystretch{1.15}
\begin{tabular}{c|c|p{5cm}<{\centering}|p{3.5cm}<{\centering}|p{3.5cm}<{\centering}}
\hline
Pre-train Dataset & Scheme & Type & Agreement (\%) $\downarrow$ & Auth Success Rate (\%) $\downarrow$ \\ \hline
\multirow{5}{*}{ImageNet} & DCmp                   & Original 2048 dimensional vector & 99.981 & 100 \\ \cline{2-5} 
                          & \multirow{4}{*}{Trans} & to 8192 dimensional vector       & 99.926 & 100 \\
                          &                        & to 512 dimensional vector        & 99.927 & 100 \\
                          &                        & to 128 dimensional vector        & 99.932 & 100 \\
                          &                        & to image                         & \textbf{98.195} & \textbf{0}   \\ \hline
\multirow{5}{*}{CIFAR-10} & DCmp                   & Original 2048 dimensional vector & 99.925 & 100 \\ \cline{2-5} 
                          & \multirow{4}{*}{Trans} & to 8192 dimensional vector       & 99.863 & 100 \\
                          &                        & to 512 dimensional vector        & 99.862 & 100 \\
                          &                        & to 128 dimensional vector        & 99.850 & 100 \\
                          &                        & to image                         & \textbf{97.330} & \textbf{0}   \\ \hline
\end{tabular}
\caption{Comparison with different verification schemes on disguised backdoor encoder. Converting feature vectors into images can effectively amplify small differences in vectors, which is beneficial for integrity verification.}
\label{compare_our}
\end{table*}

In Table \ref{compare_our}, we present the performance of different verification schemes on the backdoor encoder. Here, we select ImageNet and CIFAR-10 pre-trained backdoor encoders with STL-10 as the downstream task and execute an enhanced backdoor attack that disguises the backdoor encoder with key samples.

It is apparent that in the Transformation scheme, we assess the scenario of converting feature vectors into both high-dimensional and low-dimensional vectors, resulting in a slight decrease in agreement compared to DCmp. However, the authentication success rate remains unchanged, indicating that merely transforming feature vectors into alternative forms is insufficient to amplify the originally subtle differences. Transforming the feature vectors into the image domain yields optimal results, with an authentication success rate of 0\%.

Table \ref{precision} illustrates the authentication success rate when there is a loss of precision in the feature vectors uploaded by users. Notably, in extreme cases where only two decimal places are retained, the Trans scheme verification fails, while DCmp successfully passes the verification.

Comparing these results with the pruning rate of 10\% in Figure \ref{dis_agreement}, DCmp attains a 99.99\% agreement, with a verification success rate of 100\%, suggesting that this method lacks sufficient discriminative power for the differences in feature vectors caused by encoder changes. If the goal is to completely avoid situations where tampered encoders go undetected, Trans proves to be a superior choice, despite requiring an additional $G$ in implementation, which may make it less convenient than DCmp.

\begin{figure*}[t]
 \centering
\includegraphics[width=1\linewidth]{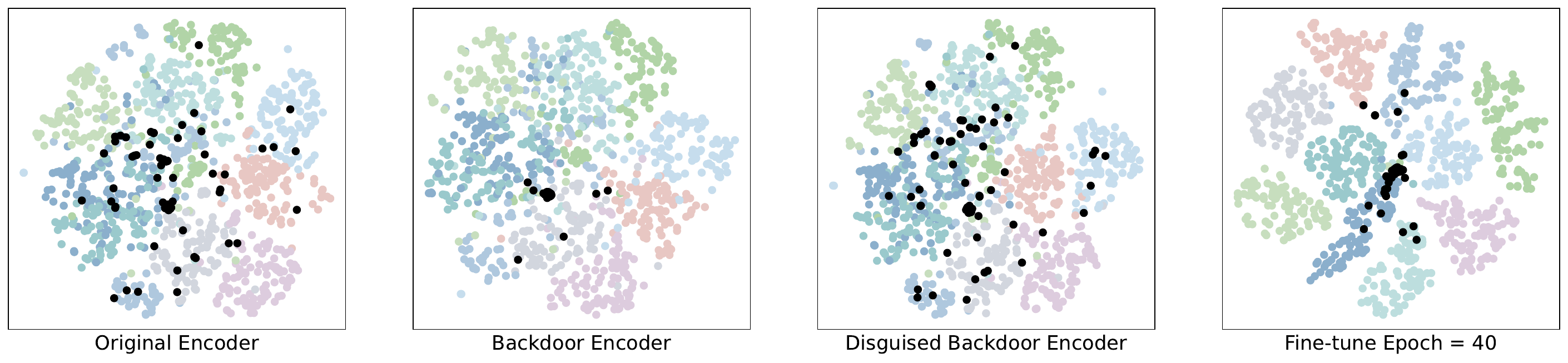}
\caption{Visualization of feature domain. With the CIFAR-10 pre-trained encoder, after fine-tuning or backdoor attacks on the encoder, the trend of OOD data changes is similar to that shown in Figure \ref{dis}.}
\end{figure*}

\end{document}